\documentclass[twocolumn,tighten]{aastex63}

\let\bfseries\mdseries

\received{XX}
\revised{XX}
\accepted{XX}
\submitjournal{AJ}

\usepackage{xcolor}
\usepackage{gensymb}

\shorttitle{WASP-107b tail}
\shortauthors{Spake et al.}

\graphicspath{{./}{figures/}}

\begin{document}

\title{The post-transit tail of WASP-107b observed at 10830\AA}

\correspondingauthor{J. J. Spake}
\email{jessica.spake@gmail.com}

\author{J. J. Spake\thanks{\textit{51 Pegasi b Fellow}}}
\affiliation{Division of Geological and Planetary Sciences, California Institute of Technology, Pasadena, CA 91125, USA}

\author{A. {Oklop{\v{c}}i{\'c}}}
\affiliation{Anton Pannekoek Institute of Astronomy, University of Amsterdam, Science Park 904, 1098 XH Amsterdam, Netherlands}

\author{L. A. Hillenbrand}
\affiliation{Department of Astronomy, California Institute of Technology, Pasadena CA 91125, USA}

\begin{abstract}

Understanding the effects of high-energy radiation and stellar winds on planetary atmospheres is vital for explaining the observed properties of close-in exoplanets.  Observations of transiting exoplanets in the triplet of metastable helium lines at 10830 \AA\, allow extended atmospheres and escape processes to be studied for individual planets.  We observed one transit of WASP-107b with NIRSPEC on Keck at 10830 \AA.  Our observations, for the first time, had significant post-transit phase coverage, and we detected excess absorption for over an hour after fourth contact. The data can be explained by a comet-like tail extending out to $\sim 7$ planet radii, which corresponds to roughly twice the Roche lobe radius of the planet. Planetary tails are expected based on 3D simulations of escaping exoplanet atmospheres, particularly those including the interaction between the escaped material and strong stellar winds, and have been previously observed at 10830 \AA\, in at least one other exoplanet. With both the largest mid-transit absorption signal and the most extended tail observed at 10830\AA, WASP-107b remains a keystone exoplanet for atmospheric escape studies.

\end{abstract}


\section{Introduction} \label{sec:intro}
Most known transiting exoplanets orbit close to their host stars and are therefore highly irradiated, which makes them likely susceptible to atmospheric escape. Population studies of bulk properties, like planet radius and orbital distance, can illuminate how mass-loss changes compositions and radii over planetary lifetimes, especially for the more numerous, smaller exoplanets \citep[e.g.][]{OwenWu2013,LopezFortney2013,Fulton2017,Ginzburg_2018}.
On the other hand, individual, gas-rich exoplanets on close orbits around bright stars are amenable to atmospheric characterization, and therefore allow atmospheric escape to be seen in action.  

Transmission spectroscopy is a powerful method for probing exoplanet atmospheres \citep[e.g.][]{SeagerSasselov2000}.  Strong absorption lines of species which are abundant at high altitudes, where gas is rarefied, are the most useful for observing atmospheric escape.  For example, the Lyman-$\alpha$ line of \textbf{ground-state} hydrogen was the first absorption line used to find exoplanet atmospheres overflowing their Roche radii, for the hot Jupiters HD209458b \citep{Vidal-Madjar2003} and HD189733b \citep{Lecavelier2010}.  \textbf{The H-alpha line, from the first excited state of hydrogen, has also been used to observe the high-altitude thermospheres of exoplanets, where mass-loss is thought to be launched \citep[e.g.][]{Jensen_2012, Yan_2018}.  Heavier, but still abundant elements, like carbon, oxygen, magnesium, and iron, have also been observed overflowing (or close to) the Roche radii of some planets (e.g. \citealt{2004ApJ...604L..69V,2013A&A...560A..54V,2019AJ....158...91S}). }

The Lyman-$\alpha$ observations were also the first to directly demonstrate ``tails" of lost material trailing behind exoplanets, \textbf{which manifest as post-transit absorption in the observed light curves.  The most dramatic example of a planetary tail is that of the hot Neptune GJ436b \citep[][]{Kulow2014,Ehrenreich2015, Lavie2017}, which blocks 56.3$\pm$3.5\% of the stellar radiation at mid-transit, and shows blue-shifted absorption over three hours after the end of the nominal one-hour transit, at radial velocities up to -100 kms$^{-1}$.  The absorption also begins two hours before the nominal ingress.  Combined, these observations imply an asymmetric cloud of material surrounding the planet, with a trailing tail shaped by stellar wind that is longer than the planet's radial distance to the star \citep{Ehrenreich2015}.  For the hot Jupiter HD189733b, excess absorption was observed for at least 30 minutes post-transit, in just one of the five observed Lyman-alpha light curves, which happened to occur 8 hours after a stellar flare \citep{Lecavelier2010,2012A&A...543L...4L}.  The result indicated variability in the escaping planetary wind, possibly caused by the varying stellar environment. For HD 209458b, which is less massive than HD189733b, strong absorption was seen at egress with high-resolution Lyman-alpha observations, which indicated a comet-like tail.  However, the observations did not have the post-transit coverage needed to observe it directly \citep{Vidal-Madjar2003}.}  Planetary tails have long been predicted (e.g. \citealt{1998ASPC..134..241S}) and studied in a number of three-dimensional (3D) numerical simulations of atmospheric escape in exoplanets \citep[e.g.][]{Bisikalo2013,Matsakos2015,  Christie2016, Bourrier2016, Carroll-Nellenback2017,  Khodachenko2019, 2019ApJ...873...89M,  Carolan2021}. 

\textbf{Pre-transit absorption has been observed in several planets, including the aforementioned GJ 436b. \cite{Fossati2010} detected an early ingress for the hot Jupiter WASP-12b in a broadband light curve at NUV wavelengths. For HD189733b, evidence of pre-transit absorption has been observed in absorption lines of three separate species: \cite{Ben_Jaffel_2013} detected absorption by CII in two separate transit events; \cite{Bourrier2013a} detected absorption by SiIII; and \cite{2015ApJ...810...13C,2016AJ....152...20C} detected absorption in H-alpha. These observations have been explained by models of a bow shock at the interface of the planetary and stellar wind, although it is unclear how much stellar variability contributes to the signal, especially as the signal size seems to correlate with stellar activity levels (e.g. \citealt{2016AJ....152...20C}). }

In recent years, the neutral helium line triplet at 10830~\AA\footnote{All absorption lines listed in this paper are at air wavelengths.} has been established as a new probe of extended and escaping exoplanet atmospheres \citep{SeagerSasselov2000, OklopcicHirata2018}. \citet{Spake2018} first observed helium in the extended atmosphere of the inflated sub-Saturn orbiting a K6 star, WASP-107b, at 10830\AA, with low-resolution spectra from the {Hubble Space Telescope/Wide Field Camera 3}. Since then there have been numerous observations of extended and escaping exoplanet atmospheres using the 10830\AA\, lines \citep[e.g.][]{Allart2018, Nortmann2018, Mansfield2018, Salz2018, Alonso-Floriano2019, Ninan2020, Palle2020, Vissapragada_2020, paragas2021metastable}, and notable non-detections, only some of which have been published \citep[e.g.][]{Kasper_2020,2020arXiv201202198Z}. Of particular relevance to this work, \cite{Nortmann2018} used CARMENES \citep{carmemes2014} to observe the 10830\AA\, transmission spectrum of WASP-69b - a warm, inflated, Saturn-mass exoplanet orbiting a K5 star.  They reported significant post-transit absorption, indicative of a planetary tail which extended slightly further than the planet's Roche radius. 

WASP-107b, which was discovered by \cite{Anderson_2017}, remains the exoplanet with the strongest helium signal observed so far.  Additionally, its retrograde, polar orbit \citep{Rubenzahl_2021}; extremely low density (0.13$^{+0.015}_{-0.013}$ gcm$^{-3}$), and distant companion (planet $c$, \citealt{2021AJ....161...70P}) have furthered interest in this unusual planet. Table \ref{tab:syspar} lists the stellar and planetary parameters adopted in this study of planet $b$. The low-resolution HST helium detection was confirmed by \citet{Allart2019}, who measured excess helium absorption of $5.54\% \pm 0.27 \%$ in the core of the deepest component of the lines, using high-resolution data obtained with the ground-based CARMENES spectrograph.  The signal is large compared to the $\sim$2\% transit depth of the planet at broad-band optical wavelengths \citep{Anderson_2017}. \cite{Kirk2020} re-observed WASP-107b at high resolution with NIRSPEC on Keck, and found a transmission spectrum consistent with the CARMENES observations. \textbf{They additionally noted excess absorption during the planet's egress, hinting at the presence of a comet-like tail. }

Hydrodynamic simulations, similar to those developed for predicting the observable signatures of atmospheric escape and planetary tails in the Lyman-$\alpha$ line, can be used to produce synthetic spectra in the He 10830 \AA\ lines \citep[e.g.][MacLeod \& Oklop\v ci\'c, in prep.]{Shaikhislamov2021, Wang2020a, wang2020metastable}. Recently, 
\citet{wang2020metastable} simulated the 10830~\AA\ observations of WASP-107b, and their fiducial model suggests the presence of a comet-like tail of material trailing behind the planet. The tail would manifest as an asymmetric transit light curve at 10830\AA, with a prolonged absorption after the fourth transit contact.  \citet{wang2020metastable} noted a hint of asymmetry in the NIRSPEC light curve based on the data originally presented by \cite{Kirk2020}. However, neither \cite{Allart2019} nor \cite{Kirk2020} observed WASP-107b with significant post-egress phase coverage, so they were unlikely to directly detect a tail.

Here we present a new transit observation of WASP-107b at high resolution around 10830\AA, with NIRSPEC on Keck.  
Our data were taken between the two transits studied in \citet{Allart2019} and \citet{Kirk2020}.
We observed over an hour's worth of high-resolution spectra after the nominal planet egress, and as a result, for the first time directly detected excess helium absorption from a comet-like tail of material lost from the planet. 
\begin{table}
    \centering
    \begin{tabular}{lccc}
    \hline
    Param. & Unit & Value & Source \\
    \hline
    $P$ & day & 5.7214742 & \cite{DaiWinn2017} \\
    $t_{c}$ & JD & 2458494.04427 & \cite{DaiWinn2017} \\
    $b$ &  & 0.07 $\pm$ 0.07 & \cite{DaiWinn2017} \\
    $i$ & $^{\circ}$ & 89.887$^{+0.074}_{-0.097}$ & \cite{DaiWinn2017}  \\
    $R_{p}/R_{*}$ & & 0.14434 $\pm$ 0.00018 & \cite{DaiWinn2017}\\
    $a/R_{\mathrm{*}}$ & & 18.164 $\pm$ 0.037 & \cite{DaiWinn2017} \\
    $M_{P}$ & $M_{\mathrm{J}}$ & 0.096 $\pm$ 0.005 &  \cite{2021AJ....161...70P} \\
    $e$ & & 0.06 $\pm$ 0.04 & \cite{2021AJ....161...70P} \\
    $\omega$ & $^{\circ}$ & 40$^{+40}_{-60}$ & \cite{2021AJ....161...70P} \\
    
    \hline
    $T_{\mathrm{eff}}$ & K & 4245 $\pm$ 70 & \cite{2021AJ....161...70P} \\
    $M_{*}$ & $M_{\odot}$ & 0.683$^{+0.017}_{-0.016}$ & \cite{2021AJ....161...70P} \\
    $R_{*}$ & $R_{\odot}$ & 0.67$\pm$0.02 & \cite{2021AJ....161...70P} \\
    
    \hline
    \end{tabular}
\caption{Adopted system parameters for WASP-107b}
\label{tab:syspar}
\end{table}

\section{Observations} \label{sec:obs}
To observe a transit of WASP-107b, we used the newly-upgraded \citep{Martin2018} NIRSPEC spectrograph \citep{mclean1998,mclean2000} on the 10-meter Keck II telescope, for one half-night. Observations were performed using NIRSPEC in its echelle mode with the N1 (Y-band) filter, having effective wavelength coverage from 0.941 to 1.122 $\mu$m, with some overlap and no gaps between the orders. The helium triplet lines appear in both Order 70 and Order 71, at rest wavelengths of 10830.34 \AA, 10830.25 \AA, and 10829.09 \AA. In this work we only present data from order 70.  We used the $0.288\arcsec  \times 12\arcsec$ slit, resulting in spectral resolution of $R= 37,000$ (corresponding to $\Delta v \approx 7.7$~km~s$^{-1}$) at the wavelength of the helium line.  

\begin{figure}
\centering
\includegraphics[width=0.48\textwidth]{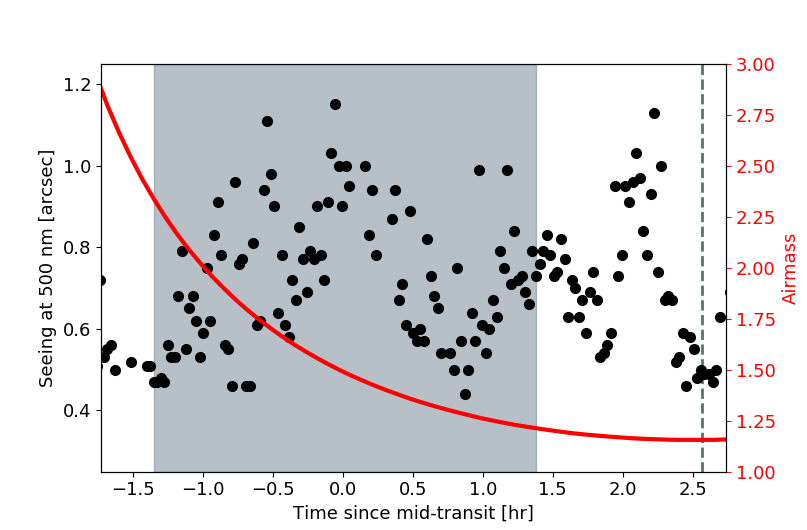}
\caption{Seeing (black dots) and airmass (red line) during our January 2019 observations. Seeing data were obtained from the Mauna Kea Weather Center.  The duration of the broadband transit is shown in gray. Dashed vertical line marks the 18-degree dawn. }
\label{fig:seeing_airmass}
\end{figure}

Our WASP-107b transit observations were performed on January 10, 2019 UT (program ID:C247, PI: Hillenbrand). The seeing was variable during the observations but averaged $\sim$0.7" (\autoref{fig:seeing_airmass}). We used a standard ABBA nodding pattern with 4$\arcsec$ throw between the nod positions and individual exposures of 300 s.  \textbf{Similarly to \cite{Kirk2020}, we used the thin blocker, which introduces fringing in Y-band data.  However, fringing effects are weak around 10830\AA, in order 70.  Additionally, because our final results were produced by finding the difference between observed spectra, the effect of fringing was subtracted out.  We did not observe any fringing pattern above the photon noise level in our final results. }

We obtained a sequence of 48 individual spectra, with SNR increasing from 55 to 80 per resolution element as the source rose in altitude. The first four spectra were taken with shorter exposures (60 s) compared to the rest of the sample (300 s), so we merged them and treated them as a single spectrum in the rest of the analysis. We discarded from our analysis two exposures (\#32 and \#42 in our sequence) taken while we were experiencing software issues, leaving 43 spectra which we used in our analysis.

On April 8, 2019 UT we observed WASP 107 at an orbital phase of 0.35 for planet $b$ -- well outside of transit.
An additional set of four spectra were again obtained with NIRSPEC's N1 filter in the ABBA nod pattern, also with the $0.288\arcsec  \times 12\arcsec$ slit, but with exposure times of 180 seconds.  Coincidentally, these extra observations occurred one night after the transit observation made by \cite{Kirk2020}.

\section{Data reduction} \label{sec:data-reduc}
First, to correct for bad pixels and cosmic rays, we applied the iterative bad pixel algorithm from REDSPEC \citep{2015ascl.soft07017K}, which we translated to Python, to every science exposure and flat field. We then made night-averaged flat fields by median-combining eleven five-second exposures of a halogen lamp, from which we subtracted the median of eleven five-second exposures taken with the lamp off, to account for detector bias and dark current.  To extract 1D spectra from the raw frames, we used a customized version of the NIRSPEC Data Reduction Pipeline\footnote{ \url{https://github.com/Keck-DataReductionPipelines/NIRSPEC-Data-Reduction-Pipeline}} (NSDRP), which is written in Python.  Before extraction, NSDRP performs flat-field correction; removal of telluric emission by pairwise subtraction of consecutive images; and spatial- and spectral rectification of misaligned images.  NSDRP had not yet been updated since NIRSPEC’s detector was upgraded, in late 2018, from an Aladdin III InSb 1024$\times$1024 array to a Teledyne HAWAII-2RG (H2RG) HgCdTe 2048$\times$2048 array.  Therefore, we updated the hard-coded, expected positions of each spectral order on the detector to their new positions, and adapted the pipeline to handle 2048$\times$2048 images rather than 1024$\times$1024 images.

To determine the wavelength solution for each 1D spectrum on the night of 2019-01-19 (``January transit" hereafter), we followed a similar method to \cite{2020arXiv201202198Z}.  First, we selected a model stellar spectrum from the PHOENIX grid appropriate for WASP-107 (T$_\mathrm{eff}$=4400K; $log(g)$=4.5; [M/H]=0.0; \citealt{2013A&A...553A...6H}) to which we added a telluric transmission spectrum\footnote{\url{https://www.gemini.edu/observing/telescopes-and-sites/sites}}, which had been shifted to account for the Earth's night-averaged radial velocity relative to the star.  We then down-sampled the resulting template spectrum to match the instrumental resolution (R=37,500).  We modelled the wavelength solution as a third-order Chebyshev polynomial (mathematically equivalent to monomials, but well-behaved around zero), and the stellar continuum as a fifth-order Chebyshev polynomial.  We then used SCIPY's differential evolution to minimize the $\chi^{2}$ between the continuum-normalised observations and the template spectrum.  This method results in wavelength-calibrated, 1D spectra in the stellar rest frame, with line positions aligning with the model to better than one pixel.

To correct for telluric absorption lines, we used ESO's \texttt{molecfit} \citep{2015A&A...576A..78K,2015A&A...576A..77S}.  \texttt{molecfit} models telluric absorption at the observatory coordinates at a given time and airmass using Global Data Assimilation System (GDAS)\footnote{\url{https://www.ncdc.noaa.gov/data-access/model-data/model-datasets/global-data-assimilation-system-gdas}} atmospheric profiles, which depend on local weather conditions.  Before inputting the spectra, we shifted them from the stellar rest frame to the barycentric rest frame, and we shifted them back to the stellar rest frame after the \texttt{molecfit} correction. Similarly to \cite{Kirk2020}, we isolated 10 telluric absorption lines across order 70 which did not overlap with strong stellar absorption lines and masked the rest of the spectra in each fit.  Due to the combined systemic and barycentric radial velocities at the time of our observations, no strong telluric absorption lines overlapped with the helium 10830\AA\, triplet, as discussed in Section \ref{sec:in-vs-out}. For continuum normalization, we used iSpec \citep{iSpec2014} to find a polynomial fit to the extracted, telluric-corrected spectra, avoiding regions with strong stellar absorption lines (such as the region containing the He triplet and the nearby Si line).

\begin{figure}
    \centering
    \includegraphics[width=0.5\textwidth]{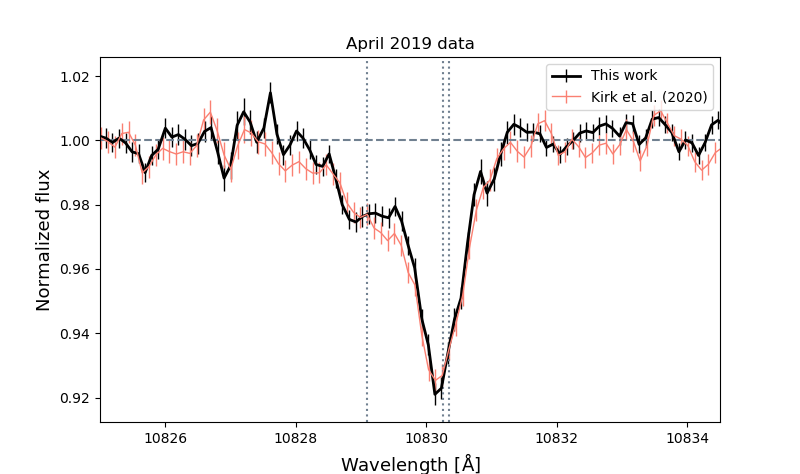}
    \caption{Excess absorption spectrum of WASP-107b (in the planet's frame) based on data obtained in April 2019 by \citet{Kirk2020}. Wavelengths of the He I triplet indicated by dotted lines. The black line shows the result obtained using our data reduction, which agrees within observational uncertainty with the result presented in \citet{Kirk2020}, shown in red. }
    \label{fig:reduction_verification}
\end{figure}

To validate our reduction pipeline we also ran it on the previously published transit observation of WASP-107b using NIRSPEC in Y-band with a wider ($0.433\arcsec  \times 12\arcsec$) slit, taken on 2019-04-06, by \cite{Kirk2020} (``April transit" hereafter).  A comparison between the resulting transmission spectrum from the April transit using our pipeline, and that published by \cite{Kirk2020} is shown in  \autoref{fig:reduction_verification}. \textbf{For both reductions, the first six spectra were used to make the out-of-transit spectrum, and and the 12 spectra between second and third contact (numbers 17 to 28 inclusive) were used to make the in-transit spectrum.}
The two independent reductions are consistent within the 1$\sigma$ error bars. For the comparison of data sets obtained using different slit widths, discussed in \S\ref{sec:in-vs-out}, we used iSpec to first degrade the spectral resolution of all extracted spectra to $R=25,000$, before combining them. 

For the extra set of four spectra of WASP-107 taken on 2019-04-08 (``April 0.35 phase" hereafter), we used the same reduction method described above, and averaged the four spectra.

\section{Results} \label{sec:results}

\subsection{January absorption spectrum}
\begin{figure}
    \centering
    \includegraphics[width=0.5\textwidth]{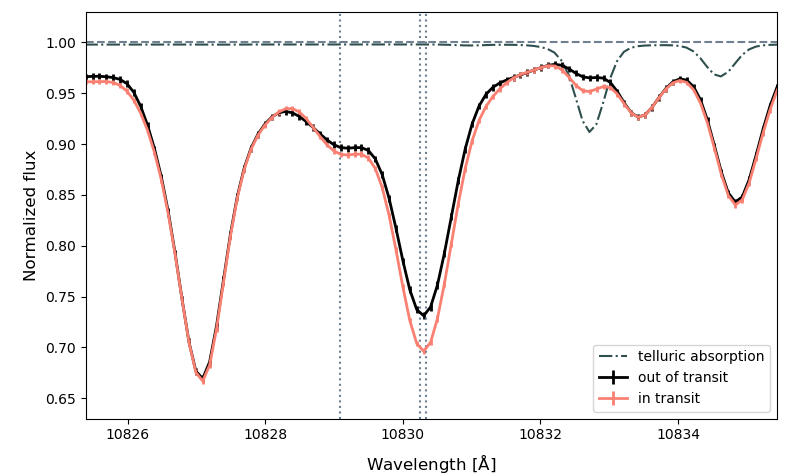}
    \includegraphics[width=0.5\textwidth]{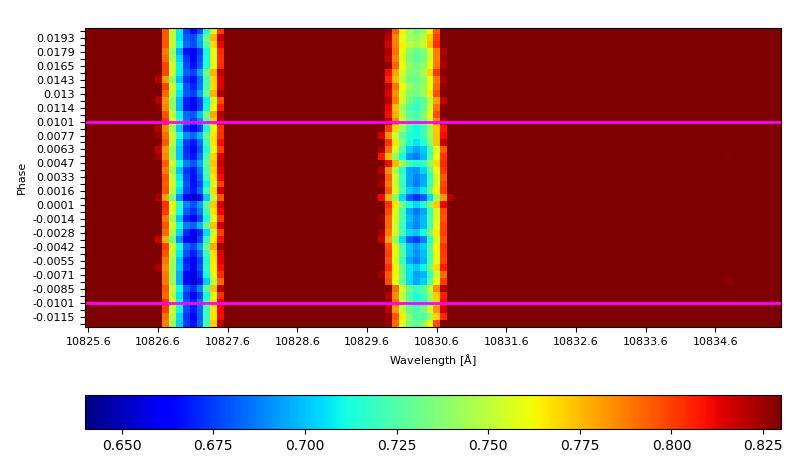}
    \caption{Top panel: in-transit and out-of-transit spectra of WASP-107, centered on the wavelengths of the He I triplet indicated by dotted lines. Excess absorption during the (broadband) transit of the planet is clearly noticeable. Bottom panel: orbital phase vs. wavelength map showing absorption depth in the He I line (the main component) and a nearby Si I line. The He I line depth increases between ingress and egress (magenta lines), while the Si I line stays constant.}
    \label{fig:in-out-spectra}
\end{figure}

For the January transit data, we made a nominal transmission spectrum for WASP-107b, assuming the planet does not have a tail.  To do this, we made an average 
``out-of-transit" spectrum, shown in \autoref{fig:in-out-spectra}, using our 12 spectra taken after the fourth contact of the nominal planet transit (after the dashed line at phase 0.0104 in \autoref{fig:lightcurve}).  We then made an average in-transit spectrum (\autoref{fig:in-out-spectra}) from our 21 spectra taken between second and third contacts of the nominal planet transit (between the \textbf{dotted} lines in \autoref{fig:lightcurve}). The ratio between the average ``in" and ``out" spectra is shown in \autoref{fig:trans-spectrm}.  The excess absorption calculated this way is dramatically less than that reported by \cite{Kirk2020}, and the peak also appears redshifted in comparison.

\subsection{Excess absorption light curves}
\label{sec:lcs}
We created an excess absorption light curve for the January transit.  To do so, we first made individual excess absorption spectra by dividing each of the 43 normalized spectra by the average ``out-of-transit" spectrum shown in Figure \ref{fig:in-out-spectra}.  Then we integrated the flux between 10829.83 and 10830.58\AA\, for each of the excess absorption spectra, shown in Figure \ref{fig:lightcurve}, top panel.  We also show the light curve from our reduction of the April transit, calculated the same way, but using the first 8 spectra in that data set to create an average out-of-transit spectrum. With this method, the January transit depth appears shallower than the April transit depth.  Also, the post-transit absorption for the January transit continues to decrease well after transit, so that the light curve reaches a level higher than the pre-transit observations at a phase of 0.02.

In the bottom panel of Figure \ref{fig:lightcurve}, we also show the same two lightcurves, added to the broadband light curve covering 9000 - 10000 \AA\, from \cite{Spake2018}, and then normalized in such a way that the transit depths match.  To do this, we calculated the weighted means of the two un-normalized light curves between the second and third contact points.  We then subtracted the difference between the means from the January transit light curve.  For normalisation, we divided both lightcurves by the weighted mean of the pre-transit April light curve (i.e. before first contact, dashed line at phase 0.0104 in Figure \ref{fig:lightcurve}).  The shape of the lightcurve is very consistent between the two observations when they are offset this way, being deeper after mid-transit, and tapering off more slowly at egress than at ingress.  Post-transit absorption is apparent in the January transit.

\subsection{In vs. Out spectra}
\label{sec:in-vs-out}
Figure \ref{fig:jan-apr-stellar} shows the averaged in-transit and ``out-of-transit" spectra for the January transit, and the equivalent spectra from our reduction of the April transit.  We stress that the January ``out" spectrum (bottom panel) was made from data taken after the nominal planetary egress, but the ``out" spectrum from the April transit was made from pre-ingress data only.  The ``out" spectra from the two nights are noticeably different, especially in the blue wings of both components of the helium triplet, where there appears to be more absorption in the January data. On the other hand, the in-transit spectra (top panel) are much more similar between the two observations.  The most significant difference in both the ``in" and ``out" spectra between the two nights occurs at wavelengths 10831.5 - 10834.0 \AA.  As we show in Figure \ref{fig:jan-apr-stellar}, those wavelengths correspond to strong telluric absorption lines which we have corrected with \texttt{molecfit}, apparently imperfectly.  Therefore we do not consider the difference at these wavelengths as astrophysical in origin, nor do we believe that telluric absorption has contaminated the helium 10830 \AA\, line triplet.

\subsection{April 0.35 phase data}
\begin{figure*}
    \centering
    \includegraphics[width=\textwidth]{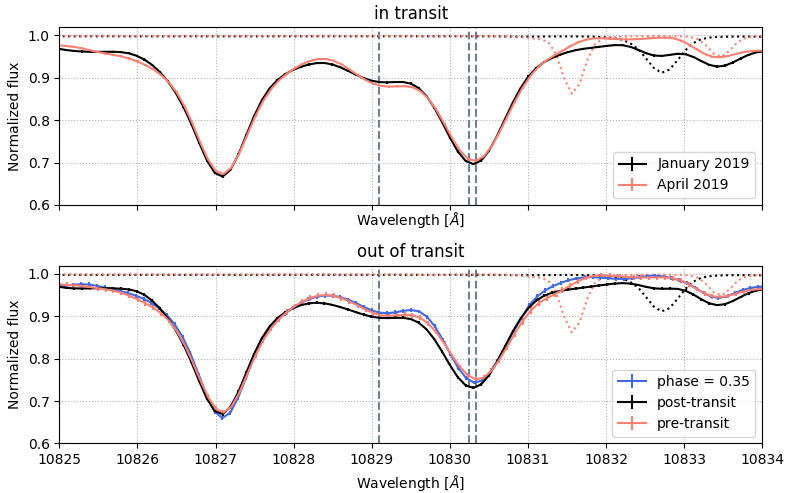}
    \caption{Comparison of in-transit and out-of-transit spectra of WASP-107 obtained with Keck/NIRSPEC in January (post-transit); April 6th (pre-transit); and April 8th (phase = 0.35) 2019. 1$\sigma$ errors are plotted as vertical bars, barely visible. For this comparison, the January and April 8th data have been degraded to spectral resolution of $R=25,000$, same as the April 6th data. Dotted lines show the telluric absorption corrections for different data sets obtained using \texttt{molecfit}. Outside of the wavelength range contaminated by telluric lines, the largest difference is seen at the wavelengths of the He I triplet (vertical dashed lines) between the spectra observed shortly after transit (black lines) and the spectra observed either shortly before (red lines) or well after transit (blue line).}
    \label{fig:jan-apr-stellar}
\end{figure*}

Figure \ref{fig:jan-apr-stellar} (bottom panel) also shows the average of our four spectra taken one night after the April transit, at an orbital phase of 0.35.  We can reasonably assume that this later spectrum is uncontaminated by transiting planetary material.  The April pre-transit spectrum is much more similar to the April 0.35 phase spectrum than to the January post-transit spectrum.  However, there are some slight differences: slightly more absorption in the pre-transit data at 10829.7 and 10831\AA, but slightly less absorption in the core of the strongest component of the line, at 10830.3\AA.

\begin{figure}
    \centering
    \includegraphics[width=\columnwidth]{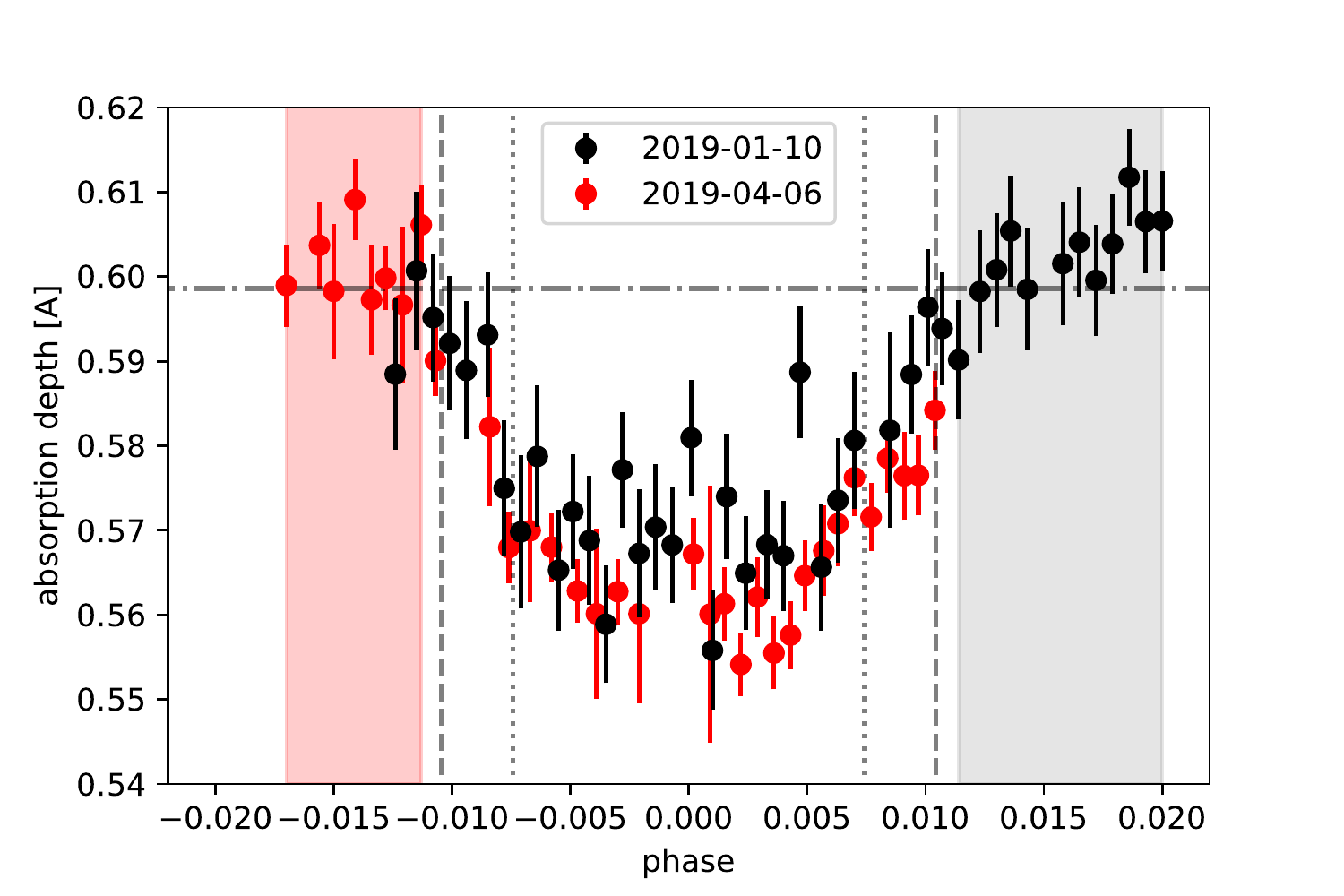}
    \includegraphics[width=\columnwidth]{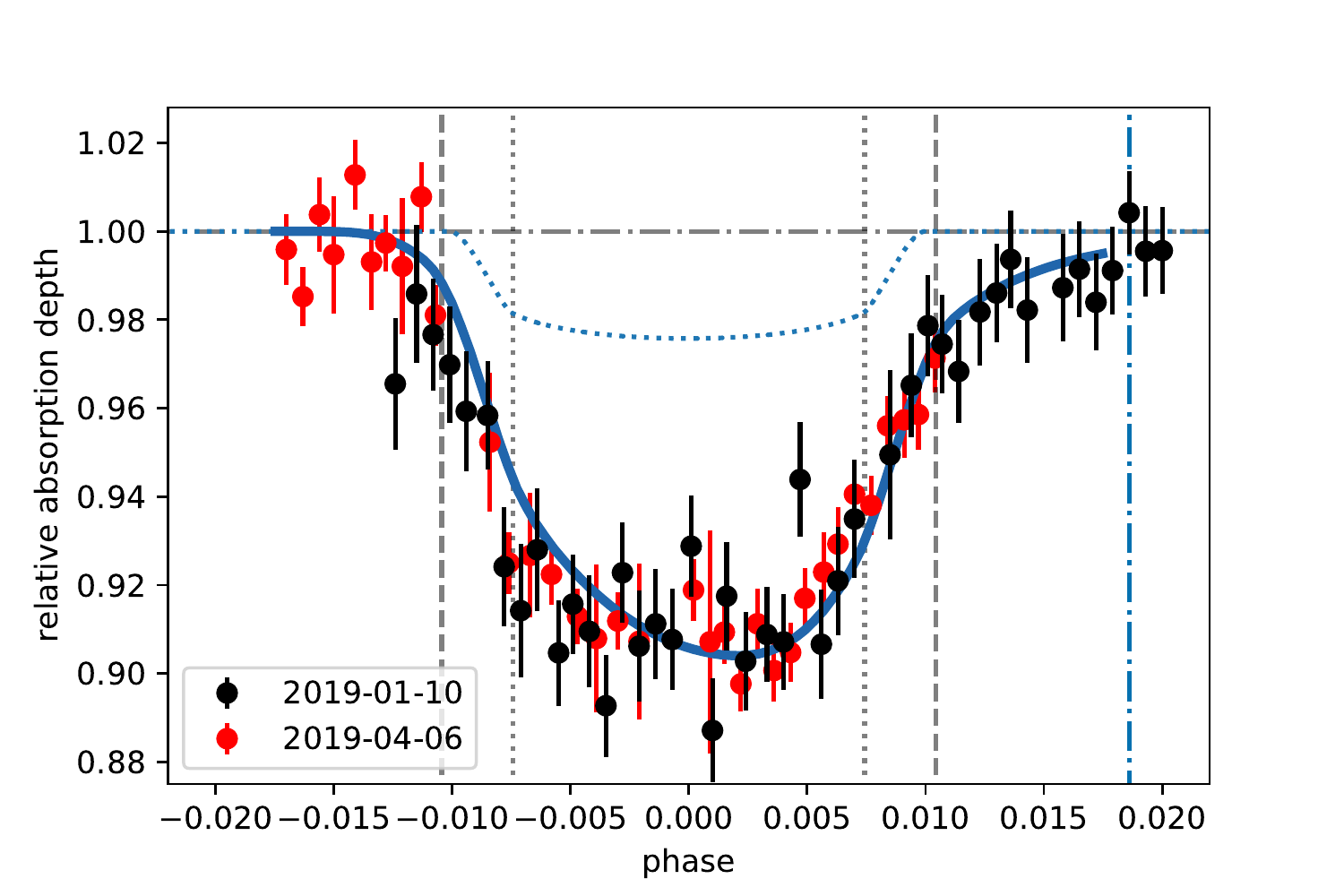}
    \caption{Top panel: Light curves showing excess helium absorption integrated over wavelengths between 1082.983 nm and 1083.058 nm. The gray(red) area shows phases used for the construction of the out-of-transit spectrum for the January(April) night. Bottom panel: same as top panel, but both light curves are added to the white-light curve from \cite{Spake2018} (dotted blue curve), and then normalized so that the transit depths match.  Solid blue curve is the light curve based on the fiducial hydrodynamic simulation from \cite{wang2020metastable}. Dash-dotted vertical line indcates the end of the post-transit absorption.}
    \label{fig:lightcurve}
\end{figure}

\begin{figure}
    \centering
    \includegraphics[width=0.5\textwidth]{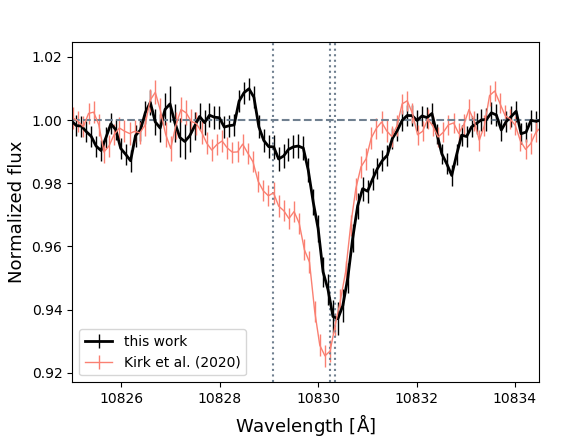}
    \caption{Absorption spectrum of WASP-107b, assuming no post-transit tail, in the reference frame of the planet observed in January 2019 (black). Wavelengths of the He I triplet indicated by dotted lines. The spectrum obtained by \citet{Kirk2020} using the same instrument in April 2019 is shown for comparison (red).}
    \label{fig:trans-spectrm}
\end{figure}

\subsection{Summary of Results}
To summarize: we detected a transit of WASP-107b at 10830 \AA, with an in-transit absorption spectrum (\autoref{fig:jan-apr-stellar}, top panel) consistent with previously published results; we demonstrated asymmetry in the transit light-curves; and we found evidence for blueshifted, post-transit absorption at 10830\AA.

\section{Analysis}\label{sec:analysis}

\subsection{Possible Causes of Variation in HeI 10830 Transmission Profiles }
Here we discuss why our January transit data and the April transit data, originally published by \cite{Kirk2020}, are different.

\subsubsection{A post-transit tail}

Because the in-transit spectra from the January and April transits are so similar (\autoref{fig:jan-apr-stellar}, top panel), the difference between the planetary transmission spectra (\autoref{fig:trans-spectrm}) must originate in the out-of-transit data (\autoref{fig:jan-apr-stellar}, bottom panel).  It is notable that our observations were taken in between the CARMENES transit presented by \cite{Allart2019}, and the Keck transit by \cite{Kirk2020}, which both had consistent absorption spectra, suggesting that the in-transit absorption profile is stable over months-long timescales within the instrumental precision. 

We find that the most likely explanation for the deeper absorption and relative blueshift in the January ``out'' spectrum is contamination from a planetary tail streaming behind the planet. Hydrodynamic simulations of escaping planetary atmospheres generally predict that the escaped material should form a comet-like tail, further shaped by the interaction with the stellar wind. For WASP-107b specifically, the modeling work of \cite{wang2020metastable} shows that the evidence of a planetary tail should be observable in the He 10830 \AA\ spectra.

\subsubsection{Contrast Effect}
The out-of-transit spectrum is set by the integrated stellar disk, but during a transit the planet blocks flux from the spatially resolved transit chord.  Therefore, wavelength-dependent differences between the integrated stellar disk and the transit chord will appear in the transmission spectrum \citep[e.g.][]{pont2008,sing2011,McCullough2014}.  
The effect of heterogeneous photospheres on exoplanet transmission spectra has been extensively modelled for FGKM stars \citep[e.g.][]{2018ApJ...866...55C,2018ApJ...853..122R,rackham2019}. \cite{2018AJ....156..189C} did the same for chromospheric lines, including He 10830\AA\,, and concluded that it was difficult to create false-positive detections of 10830\AA\, in exoplanet atmospheres with the contrast effect.

However, stellar rotation combined with the oblique orbit of WASP-107b \citep{Dai_2017,Rubenzahl_2021} can result in the planet passing over different quiet or active regions of the star at different epochs.  Therefore we could expect the contrast effect, and hence the measured transmission spectrum, to vary from visit to visit for this system. In principle, this could cause the variations we see in the January vs April WASP-107b transit spectra (\autoref{fig:trans-spectrm}). Such visit-to-visit variations have been observed in 10830\AA\, transmission spectra of HD189733b by \cite{2020A&A...639A..49G}, who analysed 5 transits of the planet and saw changes in the transmission spectrum which they attributed to the contrast effect. However, for HD189733b, there is strong evidence for spot- and faculae crossing events in the time-series light curves from the visits which deviate from the average transmission spectrum. There is little evidence for similar spot or faculae crossing events in the January transit light curve of WASP-107b at 10830\AA, and the shape of the January and April lightcurves presented here are remarkably similar when the transit depths are forced to match.

Finally, the contrast effect cannot explain why we see a significant difference in the ``out-of-transit'' spectrum, while the in-transit spectra look more similar (\autoref{fig:jan-apr-stellar}).

\subsubsection{Variability in the Planetary Wind}
The \textbf{mid-transit absorption by} the planet itself may vary from epoch to epoch - either due to a changing mass loss rate, or a changing population fraction of the excited 2$^3$S (metastable) state of helium, which is the origin of the 10830~\AA\ absorption line.  Such changes could be caused by varying extreme-ultraviolet (EUV) flux from the star \citep[e.g.][]{Oklopcic2019}, which is dependent on the level of magnetic activity of the planet-facing hemisphere.  For example, a higher stellar EUV flux during the April transit could lead to a greater \textbf{mid-transit} absorption signal compared to that of the January transit (as seen in Figure \ref{fig:trans-spectrm}). However, we might expect to see this in the out-of-transit data as increased stellar 10830~\AA\, absorption, since the stellar EUV flux also populates the metastable state in the stellar chromosphere.  In fact, the April out-of-transit  10830\AA\, absorption is weaker than that of January (assuming no planetary tail). 

Several studies, \textbf{including \cite{wang2020metastable} and \cite{2019ApJ...873...89M},} have found that shear instabilities due to interaction with a stellar wind can change the level of absorption observed \textbf{in transit light curves of escaping atmospheres. The greatest variability is seen in the pre- and post-transit absorption, however, they also report low levels of variability between second and third contact.}  For the fiducial model of \cite{wang2020metastable}, which best-fits the observations of \cite{Allart2019} and \cite{Kirk2020}, \textbf{the mid-transit} variations are small, totalling less than 5\% of the equivalent width of the absorption line.  The consistency of the \cite{Allart2019} and \cite{Kirk2020} transmission spectra, which sandwiched our observations in time, also argues against significant variations in the planetary wind.  In any case, all of the \cite{wang2020metastable} models which include a stellar wind suggest that the planet should have a trailing tail---which would contaminate the post-transit spectra---so it is difficult to avoid a tail as the cause of the observed variation.

\subsection{Modelling the Outflowing Gas from WASP-107b}
Because there is no significant pre-transit baseline on the night of our January transit observations, we cannot conclusively measure when the excess absorption diminishes entirely from the light curve---and therefore measure the size of an extended tail.  However, we can make an estimate, if we assume that the transit depth at 10830\AA\, did not change between January and April (which may not necessarily be true because of the reasons discussed above).  When the transit depths are matched, as described in Section \ref{sec:lcs}, the January transit light curve reaches the unitary baseline at a phase of around 0.0186 (dash-dotted line in Figure \ref{fig:lightcurve}).  This phase corresponds to an orbital length of 5.0$\times 10^{5}$ km; $\sim$7.0 planet radii; or roughly twice the planetary Roche lobe radius (calculated using the parameters in Table \ref{tab:syspar} and Equation 2 of \citealt{1983ApJ...268..368E}).

The effect of stellar wind on escaping exoplanet atmospheres has been a topic of considerable interest in recent years \citep[e.g.][]{StoneProga2009, Matsakos2015, Christie2016, Villarreal2018, Carolan2021}. For example, \cite{2019ApJ...873...89M} used hydrodynamical models of escaping atmospheres to study the effects of varying stellar environments---including stellar winds---in the context of Lyman-alpha transit observations of irradiated gas giants.  Their results, which concerned the bulk loss of atmospheric material, are relevant to WASP-107b, even though we observed metastable helium rather than hydrogen.  They showed that with a very weak stellar wind, lost material from a highly-irradiated planet forms a low-density torus around the star.  This would not impart a significant time-varying signal on observed spectra.  On the other hand, the radial force of a strong stellar wind directs the lost material into a dense stream (which is also shaped by the rotational and orbital motion of the planet).  It is this radially-directed stream which results in excess blue-shifted absorption post transit.   It may therefore be possible to infer the relative strengths of the planetary and stellar winds from observations of post-transit tails like that of WASP-107b (MacLeod \& Oklop\v ci\'c, in prep.).

In Figure \ref{fig:lightcurve} we show the fiducial light curve model of WASP-107b at 10830\AA\, from the hydrodynamical simulations of \cite{wang2020metastable}, which include stellar wind effects. \textbf{They used a stellar wind density at WASP-107b's orbital distance of 1.7$\times 10^{-19}$ g cm$^{-3}$ (about 20 times the solar value) and high-energy fluxes slightly higher than a typical K5 star, and found a mass-loss rate of around 1.02 $\times 10^{-9}$ Earth-masses per year.}  We did not fit the model to our data, yet the observed post-transit absorption is remarkably similar to that of the simulation, as is the overall shape of the light curve.  \textbf{The model is slightly shallower than the data from 1st contact till mid-transit.  \cite{2019ApJ...873...89M} show how increasing the stellar wind density can compress the mass-loss flow on the eastern limb of the planet, and thus truncate any pre-transit absorption signals. Perhaps these observations might therefore be better explained by a more rarefied stellar wind.}

\textbf{Because our observations provide the only post-transit coverage of WASP-107b at high resolution around 10830\AA\, so far, it is difficult to search for variability caused by shear instabilities, which is expected to be most dramatic in the pre- and post-transit epochs.  There is a small amount of pre-transit absorption in the light curves of \cite{Allart2019}, \cite{Kirk2020}, and this work.  However, all three data sets are consistent within the 1$\sigma$ errors when we normalise our lightcurve as described in Section \ref{sec:lcs}.  Because we normalised the lightcurve by matching our data to the \cite{Kirk2020} between the second and third contacts only, our analysis would not be affected by pre- or post-transit variability.  Finally, our main conclusion - that WASP-107b has a long, post-transit tail - would be unaffected by incorrect light curve normalisation, because the normalisation would not change the prolonged egress that we observed. }

\section{Discussion and conclusions}
To summarize, we directly observed excess, post-transit absorption for WASP-107b at 10830\AA, with NIRSPEC on Keck.  The most likely cause is a comet-like tail of lost atmospheric material trailing the planet, which is a common feature seen in 3D numerical simulations of escaping exoplanet atmospheres.  The post-transit absorption we measured is remarkably similar to the simulations of WASP-107b from \cite{wang2020metastable}.  \textbf{Hints of a tail were noted by \cite{Kirk2020}, who observed excess absorption during egress, however}, they did not have as much post-transit phase coverage as we did.  Our transmission spectrum, constructed assuming the nominal `out-of-transit' epochs (i.e. assuming no excess helium absorption in the post-egress spectra) is significantly different than the two previously published absorption spectra at 10830\AA, which were constructed using out-of-transit spectra derived from either solely \citep{Kirk2020} or mostly \citep{Allart2019} pre-transit epochs.  We cannot rule out contribution from the contrast effect, i.e. crossing over different active/quiet regions of the star from epoch to epoch.  However, WASP-107b's asymmetrical lightcurve and a deeper and more blueshifted post-transit spectrum (compared to previous observations)
point to a tail as the most likely physical cause of the difference.

There are two other exoplanets which show evidence for post-transit tails at 10830\AA: the warm, Saturn-mass WASP-69b \citep[][]{Nortmann2018} and the hot Jupiter HD189733b \citep[e.g.][]{2020A&A...639A..49G}.  The post-transit absorption for HD189733b is of fairly low significance, and therefore \cite{2020A&A...639A..49G} do not provide an estimate of its extent.  For WASP-69b, \cite{Nortmann2018} measured significant post-transit absorption for at least 20 minutes after fourth contact in two transit observations, and they also observed a blueshift in the 10830\AA\, lines at that time.  They calculated that WASP-69b's tail extends to around 1.7 $\times 10^{5}$ km; 2.4 Jupiter radii; or 1.2 times WASP-69b's Roche lobe radius.  WASP-69b is a particularly interesting planet to compare to WASP-107b, because they both have fairly similar stellar host types (K6/K5, respectively); orbital distances (0.55/0.45 AU); and planetary radii (0.95/1.06 Jupiter radii), but quite different masses (0.096/0.26 Jupiter masses).  At 5.0$\times 10^{5}$ km, WASP-107b appears to have a much more extended post-transit tail at 10830\AA, which is perhaps unsurprising, since WASP-107b has much deeper overall excess absorption at mid transit (6\% vs 3\%). Perhaps it is the lower gravitational potential of the planet which leads to a more extended cloud of material encircling and trailing the planet.  In any case, WASP-107b appears to have the most dramatic post-transit tail of any exoplanet observed at 10830\AA\ so far.

WASP-107b has a nominal planetary transit duration of 2.7 hours, plus a tail which shows excess absorption for around 80 minutes after the nominal egress.  \textbf{A full transit, with pre-egress and a post-tail baseline will be necessary to better constrain the size and kinematics of the tail.  In order to observe the entire transit from the ground, with ample pre- and post- transit baseline, the mid-transit time needs to occur close to the middle of a given night.  One half-night of telescope time, we have shown, is not enough.  Because it only transits once every 5.7 days, such events will be rarer for WASP-107b than for typical hot Jupiter exoplanets, which usually have shorter orbital periods.}
A scheduled transit observation lasting 6.5 hours (program ID:1224, PI:Birkmann) with NIRSpec on the James Webb Space Telescope (JWST), which covers 10830\AA, may be long enough to cover the required phases \textbf{to observe the full tail}.  The medium resolution of the JWST observations (R$\sim$2,700) will impede detailed characterisation of the helium absorption line profile, but the precision of the spectro-photometric lightcurve will be unmatched.  Since WASP-107b remains the planet with the largest absorption depth and signal-to-noise ratio, and now has the largest observed tail at 10830.3\AA, observations of a full transit of the planet and tail will be essential for validating hydrodynamical models and investigating the influence of stellar wind on escaping atmospheres. Finally, we advise caution in selecting spectra to use for the baseline stellar flux for any transmission spectrum, when the spatial extent of the planetary atmosphere is unknown.

\facility{Keck:II (NIRSPEC)} 
\software{SciPy \citep{2020SciPy-NMeth};  NumPy \citep{van2011numpy};  matplotlib \citep{Hunter:2007}; Astropy \citep{2013A&A...558A..33A}, iSpec \citep{iSpec2014}}

\acknowledgements
We are extremely grateful to Carlos Alvarez and Greg Doppmann for their expert assistance with Keck/NIRSPEC operations, 
especially during the early post-upgrade epoch of our observations, 
and to Emily Martin for helpful pre-observation conversations concerning the upgraded instrument.
We thank Trevor David, James Kirk, Ryan Rubenzahl, Michael Zhang, Fei Dai, Heather Knutson, and Romain Allart for helpful discussions.  We also thank Lile Wang and Fei Dai for their model light curve.

We thank the expert referee for helpful comments.

AO gratefully acknowledges support from the Dutch Research Council NWO Veni grant.  JJS is supported by the Heising Simons \textit{51 Pegasi b} postdoctoral fellowship.

The data presented herein were obtained at the W. M. Keck Observatory, which is operated as a scientific partnership among the California Institute of Technology, the University of California and the National Aeronautics and Space Administration. The Observatory was made possible by the generous financial support of the W. M. Keck Foundation. The authors wish to recognize and acknowledge the very significant cultural role and reverence that the summit of Maunakea has always had within the indigenous Hawaiian community. We were most fortunate to have the opportunity to conduct observations from this mountain.

\bibliography{sample63}



\end{document}